\documentclass[aps,prc,twocolumn,showkeys,showpacs]{revtex4}

\usepackage{amssymb}
\usepackage{graphicx}% Include figure files
\usepackage{dcolumn}% Align table columns on decimal point
\usepackage{bm}% bold math
\usepackage{color}
\newcommand{\req}[1]{(\ref{#1})}
\newcommand{\bel}[1]{\begin{equation}\label{#1}}
\newcommand{\belar}[1]{\begin{eqnarray}\label{#1}}
\def\mev{\;{\rm MeV}}

\def\eps{\epsilon}

\begin{document}

\title{Fission dynamics at low excitation energy}

\author{Y.~Aritomo$^{1}$, S.~Chiba$^{1}$ and F.~Ivanyuk$^{1,2}$}

\affiliation{$^{1}$ Research Laboratory for Nuclear Reactors, Tokyo Institute of Technology,
Ookayama, Meguro-ku, Tokyo, 152-8850, Japan}%
\affiliation{$^{2}$ Institute for Nuclear Research, Kiev, Ukraine}%
%\affiliation{$^{1}$ Japan Atomic Energy Agency, Tokai, Ibaraki, 319-1195, Japan}%
%\affiliation{$^{2}$ Flerov Laboratory of Nuclear Reactions, JINR, Dubna, 141980, Russia}%
%\affiliation{$^{2}$ Department of Physics, Tohoku University, Sendai 980-8578, Japan}%

%\author{Y.~Aritomo}
%\affiliation{Flerov Laboratory of Nuclear Reactions, JINR, Dubna, Russia}%

%\affiliation{Japan Atomic Energy Agency, Tokai, Ibaraki, 319-1195, Japan}%


\begin{abstract}

The mass asymmetry in the fission of $^{236}$U at low
excitation energy is clarified by the analysis of the
trajectories obtained by solving the Langevin equations for the
shape degrees of freedom. It is demonstrated that the position of
the peaks in the mass distribution of fission fragments is
determined mainly by the saddle point configuration originating from the
shell correction energy. The width of the peaks, on the other
hand, results from the shape fluctuations close to the scission point caused by the random
force in the Langevin equation. We
have found out that the fluctuations between elongated and
compact shapes are essential for the fission process. According
to our results the fission does not occur with continuous
stretching in the prolate direction, similarly to that observed
in starch syrup, but is accompanied by the fluctuations between
elongated and compact shapes. This picture presents a new
viewpoint of fission dynamics and the splitting mechanism.

\end{abstract}

\pacs{25.70.Jj, 25.85.w, 27.90.+b, 28.41.-i}
\keywords{fission process, two-center shell model, Langevin equations, dynamical trajectories, mass distribution and total kinetic energy of fission fragments}

%\documentclass{elsart}
%\usepackage{natbib}
%\usepackage{graphicx}
%%\renewcommand{\baselinestretch}{1.65}
%%\usepackage[figuresright]{rotating}
%\begin{document}
%\runauthor{Y.~Aritomo}
%\begin{frontmatter}
%\title{Analysis of dynamical process using mass distribution of fission fragments in heavy-ion reactions }

%\begin{document}

\maketitle

% main text

%%%%%%%%%%%%%%%%%%%%%%%%%%%%%%%%%%%%%%%%%%%%
%% MAINMATTER   Fig.~\ref{fig_mass36s}
%%%%%%%%%%%%%%%%%%%%%%%%%%%%%%%%%%%%%%%%%%%%

\section{Introduction}

Since the  discovery of fission of uranium in 1938 \cite{harn39,meit39},
the principle of this phenomenon has been studied owing to its scientific interest.
The application of fission process to the supply of power was realized soon after its discovery.
However, the mass-asymmetric fission remained a puzzle as far as nuclei were described in the analogy with
the liquid drop \cite{bohr39}.
The origin of the asymmetry in the mass distribution of fission fragments (MDFFs) nowadays is related to the shell structure of the fissioning nucleus.
Many theoretical dynamical models have been applied to  nuclear fission at low excitations in an attempt to explain its mechanism \cite{schm03,schm03a,asan04,gout05,rand11,ivan13}.

In order  to investigate the time evolution of the nuclear shape during the
fission process a dynamical approach using the Langevin equation can be used.
 In our previous study \cite{arit13} this approach was
applied to the fission of $^{234, 236}$U and $^{240}$Pu at low
excitation energies with account of the shell structure of these
nuclei. In these calculations we obtained an asymmetric MDFF and
the total kinetic energy (TKE) of the fission fragments that
agreed well with the experimental data.

In the present work, we attempt to clarify the origin
of the asymmetric MDFF of $^{236}$U at a low excitation energy by the analysis of the time evolution of nuclear shape and trajectories calculated within the Langevin approach.
We have found the factors determining the positions and widths of the peaks in the MDFF:
the former is mainly related to the positions of the fission saddle, which is influenced
by the shell correction
energy, and the latter is related to the thermal fluctuation caused by the random force in the Langevin equation close to the scission point.

In addition, we observed a new phenomenon in the mechanism of
fission dynamics: the fluctuation of the shape of fissioning
nucleus between the compact and elongated  configurations on the
way to the scission point. By comparing the TKE of the fission
fragments obtained experimentally and by our calculation, we
confirmed that such a configuration is realized there. This leads
to the picture in which the fission does not occur in the manner
of starch syrup, which grows with continuous stretching until the
neck radius gets very small.

Below we present the arguments in favor of this new interpretation of fission dynamics at low excitation energy.
The paper is organized as follows.
In Sec.~II, we describe in detail the framework of the model.
In Sec.~III, we discuss the potential energy landscape and  reveal the effect of the shell structure on the mass-asymmetric fission of $^{236}$U at $E^{*}=20$ MeV by analyzing the dynamical trajectories.
In Sec.~IV, we investigate the configuration at the scission point.
%The role of the friction tensor in fission dynamics is described in Sec.~V.
In Sec.~V, the TKE of the fission fragments is discussed.
A short summary of this study and further discussion are presented in Sec.~VI.
%%%%%%%%%%%%%%%%%%%%%%%%%%%%%%%%%%%%%%%%%%%%%%%%%%%%%%%%%%%%%%%%%%%%%%%%%%%%%%%%%%%%%%%%%%%
\section{The model}
%%%%%%%%%%%%%%%%%%%%%%%%%%%%%%%%%%%%%%%%%%%%%%%%%%%%%%%%%%%%%%%%%%%%%%%%%%%%%%%%%%%%%%%%%%
We use the fluctuation-dissipation model and
employ the Langevin equations \cite{arit04} to investigate the dynamics of the fission process.
The nuclear mean field is defined by the two-center shell model potential
\cite{maru72,sato78} which includes the central part, ${\bf ls}$ and ${\bf l}^2$ terms. The central part consists of two oscillator potentials smoothly joined by the fourth order polynomial.
%, see Fig. \ref{def1}.
Within the two-center shell-model parameterization (TCSMP) the
shape is characterized by 5 deformation parameters: the distance
$z_0$ between the centers of left and right oscillator potentials,
the deformations $\delta_1$ and $\delta_2$ of the left and right
oscillator potentials, the neck parameter $\eps$ and the mass
asymmetry $\alpha=(A_{1}-A_{2})/(A_{1}+A_{2})$, where $A_{1}$ and
$A_{2}$ denote the mass numbers of heavy and light fragments
\cite{arit04} (in case of the shape separated into two fragments)
or the masses of the right and left parts of the compact nucleus.
Please, note, that within TCSMP the shape is divided in parts by
the point $z=0$.

The formal definition of these parameters is given in Appendix and demonstrated in Fig.~\ref{def1}.

In order to reduce the computation time we use in this work the
restricted deformation space. We assume that the
parameters $\delta_1$ and $\delta_2$ are the same,
$\delta_1=\delta_2=\delta$. The parameters $\delta_1$ and
$\delta_2$ fix the deformation of {\it potential} in "outer"
region, namely for $z\leq z_1$ or $z_2\leq z$, see Fig.
\ref{def1}. The deformation of {\it fragments} depends not only
on $\delta_1$ and $\delta_2$ but on all other parameters, $z_0,
\epsilon$ and $\alpha$. Changing elongation or mass asymmetry one
can get the fragments with different deformation even if
$\delta_1$ is put equal to $\delta_2$.

The neck parameter $\epsilon$ is kept fixed. Here, like in the previous works, we use the value
$\epsilon=0.35$, which was recommended in Ref. \cite{yama88} for the fission process.
Keeping $\eps$ fixed does not mean that the neck radius is fixed.
It is clearly demonstrated by Fig. \ref{def1} in the Appendix.
The neck radius in TCSMP depends on
all deformation parameters. When we change the elongation, mass asymmetry or deformation of fragments the neck radius changes too.

Thus, with three deformation parameters $z_0, \alpha$ and $\delta$ we take into account the three most important degrees of freedom for the fission process: elongation, mass
asymmetry and the neck radius.

For the sake of convenience, like in previous works, instead of
$z_0$ we employ the scaled coordinate  $\bar z_0$ defined as
$\bar z_0=z_{0}/(R_{CN}B)$, where $R_{CN}$ denotes the radius of
a spherical compound nucleus and $B$ is defined as:
$B=(3+\delta)/(3-2\delta)$. The three deformation parameters
$\bar z_0, \alpha$ and $\delta$ are considered as the dynamical
variables.

These three collective coordinates may be abbreviated as $q$, with $q=\{\bar z_0,\alpha,\delta\}$.
For a given value of the intrinsic excitation energy characterized by the temperature $T$
the potential energy is defined as the sum of the liquid-drop (LD) part and the microscopic (SH) part,
\belar{vt1}
V(q, T)&=&V_{\rm LD}(q)+V_{\rm SH}(q, T),\nonumber\\
V_{\rm LD}(q)&=&E_{\rm S}(q)+E_{\rm C}(q),\,\nonumber\\
V_{\rm SH}(q, T)&=&[\Delta E_{shell}(q)+\Delta E_{pair}(q)]\Phi (T),\nonumber\\
\Phi (T)&=&\exp ({-\bar a T^2}/{E_{\rm d}} ).
\end{eqnarray}
In \req{vt1} the $V_{\rm LD}$ is the potential energy calculated with the
finite-range liquid drop model \cite{krap79}, given by the sum of the surface
energy $E_{\rm S}$ and the Coulomb energy $E_{\rm C}$.
The microscopic energy $V_{\rm SH}$  at $T=0$ is calculated as the sum of the shell correction energy $\Delta E_{shell}$ and the pairing correlation correction energy $\Delta E_{pair}$. We assume that the angular momentum of fissioning nucleus at the low excitation energy is not large, so the rotational energy is not included in \req{vt1}.

The $\Delta E_{shell}$ is calculated by Strutinsky method \cite{stru67,brdapa} from the single-particle levels of the two-center shell model potential
\cite{maru72,suek74,iwam76} as the difference between the sum of single-particle energies
of occupied states and the averaged quantity.

The $E_{pair}$ was evaluated in BCS approximation following
\cite{nilsson69,brdapa}. The averaged part of the pairing
correlation energy was calculated assuming that the density of
single-particle states is constant over the pairing window. The
pairing strength constant was related to the average gap
parameter $\widetilde\Delta$ by solving the gap equation in the
same approximation and adopting for $\widetilde\Delta$ the value
$\widetilde\Delta=12 /\sqrt{A}$ suggested in  \cite{nilsson69} by
considering the empirical results for the odd-even mass
difference.

The  $E_{\rm d}$ in (\ref{vt1}) is the  damping parameter of the
shell correction chosen to be equal to 20 MeV like in
\cite{igna75}.
In the level density parameter \cite{toke81} both the shell effects \cite{igna75,igna79} and the dependence of average part $\widetilde{a}$ on the deformation were taken into account,
\begin{eqnarray}\label{denslevpar}
\bar a&=&
%\frac{A}{14.61} \left (1+3.114 A^{-1/3}+5.626 A^{-2/3} \right ),
\left\{1+\frac{V_{SH}(T=0)}{E_{int}}\left[1-\exp{\left(-\frac{E_{int}}{E_d}\right)}\right]\right\}\widetilde{a}(q),\nonumber\\
\widetilde{a}(q)&=&a_1A + a_2A^{2/3}B_s(q),
\end{eqnarray}
with $A$ being the mass number of fissioning nucleus and $B_s$ - the reduced surface energy, see \cite{karp03}. The $E_{int}$ in \req{denslevpar} is the intrinsic excitation energy, see \req{eint} below, calculated at each step of integration of equations of motion.

\begin{figure}[ht]
\includegraphics[width=.4\textwidth]{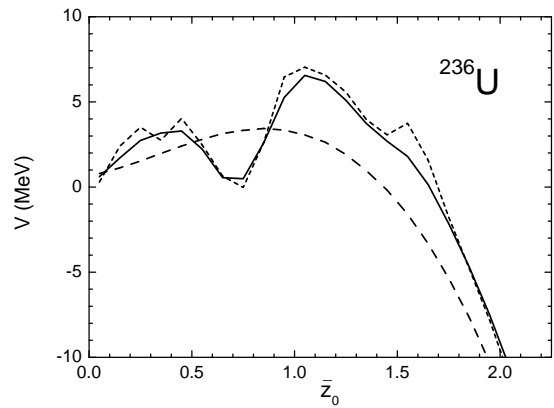}
\caption{Deformation dependence of the potential energy of
$^{236}$U at $T=0$ calculated by TWOCTR  with fixed $\delta=0,
\alpha=0$ and $\epsilon=0.35$. The $V_{\rm LD}$ and $V_{\rm
LD}+V_{\rm SH}$ are denoted by the dash and solid lines,
respectively. The short-dash line shows the $V_{\rm LD}+\Delta
E_{shell}$. } \label{fig1}
\end{figure}

To calculate the potential energy,  we employed the
macroscopic-microscopic method and TWOCTR code of the two-center
shell model \cite{suek74,iwam76,sato79}. In this code, the
parameters of the finite-range liquid drop model \cite{krap79}
are used  $r_{0}=1.20$ fm, $a=0.65$ fm, $a_{s}=21.836$ MeV and
$\kappa_{s}=3.48$, where $r_{0}$ and $a$ are the nuclear-radius
constant and the range of the Yukawa folding function, $a_{s}$
and $\kappa_{s}$ are the surface energy constant and the
surface-asymmetry constant, respectively. The
potential  energy $V_{\rm LD}$ and $V_{\rm LD}+V_{\rm SH}$
(denoted by the dash and solid lines, respectively) for $^{236}$U
with $\delta=0, \alpha=0$ and $\epsilon=0.35$, calculated by
TWOCTR is presented in Fig.~\ref{fig1}.
%%%%%%%%%%%%%%%%%%%%%%%%%%%%%%%%%%%%%%%%%%%%%%%%%%%%%%%%%%%%%%%%%%%%%%%%%%%%%%%%%%%%%%%%%%
%%%%%%%%%%%%%%%%%%%%%%%%%%%%%%%%%%%%%%%%%%%%%%%%%%%%%%%%%%%%%%%%%%%%%%%%%%%%%%%%%%%%%%%%%%

We assume that the temperature dependence of the microscopic
energy $V_{\rm SH}$ is expressed by the factor $\Phi (T)$ in
Eq.(\ref{vt1}). This dependence was suggested a long ago
\cite{asan04} and was confirmed by many years of experience. We are
aware that temperature dependence of $\Delta E_{shell}$ and
$\Delta E_{pair}$ is not the same, see also \cite{gaim91,schm82}.
Within the BCS approximation the pairing correlations disappear
above critical temperature $T_{crit}\approx 0.5 \div 0.6 \mev$
while the $\Delta E_{shell}$ becomes negligible small at
$T\approx 2 \mev$. In present calculations at $E^{*}=20 \mev$ in
most cases the local temperature is larger than the $T_{crit}$.
However, even at $T=0$ the $\Delta E_{pair}$ is small
compared with $\Delta E_{shell}$, see Fig. \ref{fig1}
(the short-dash line shows the $V_{\rm LD}+\Delta
E_{shell}$). So, the
use of approximation \req{vt1} should not lead to a large inaccuracy
of calculated results.

In our previous study \cite{arit13}, we discussed the temperature dependence of the shell correction energy and the effect of this dependence on
the fission process and the MDFF for $^{236}$U at $E^{*}=20$ MeV.
Using the several values of the shell damping energy, we investigated the affection of the MDFF.
We have found out that the gross features of MDFF did not change so much in this system.

The multidimensional Langevin equations \cite{arit04} are given as
\begin{eqnarray}
\frac{dq_{i}}{dt}&=&\left(m^{-1}\right)_{ij}p_{j},\nonumber \\
\frac{dp_{i}}{dt}&=&K_{i}
                 -\frac{1}{2}\frac{\partial}{\partial q_{i}}
                   \left(m^{-1}\right)_{jk}p_{j}p_{k}
-\gamma_{ij}\left(m^{-1}\right)_{jk}p_{k}\nonumber \\
                  &+&g_{ij}R_{j}(t),
\label{lange1}
\end{eqnarray}
where $q_i = \{\bar z_0, \delta, \alpha\}$ and
%$p_{i} = dq_{i}/dt$
$p_{i} = m_{ij} dq_{j}/dt$
is a momentum conjugate to coordinate $q_i$.
The summation is performed over repeated indices.
The conservative force in \req{lange1} is represented by the derivative of free energy with respect to deformation, $K_i\equiv -{\partial F}/{\partial q_{i}}$, with $F(q, T)=V(q, T) - \bar a T^2$.

The $m_{ij}$ and $\gamma_{ij}$ in \req{lange1} are the shape-dependent collective inertia and the friction tensors, respectively.
The wall-and-window one-body dissipation
\cite{bloc78,nix84,feld87} is adopted for the friction tensor which can describe the
pre-scission neutron multiplicities and total kinetic energy of fragments simultaneously \cite{wada93}.
A hydrodynamical inertia tensor with the Werner-Wheeler approximation
for the velocity field \cite{davi76} was used here.

The normalized random force $R_{i}(t)$ is assumed to be that of white noise, {\it i.e.},
\bel{noise}
\langle R_{i}(t) \rangle=0, \qquad\langle R_{i}(t_{1})R_{j}(t_{2})
\rangle = 2 \delta_{ij}\delta(t_{1}-t_{2}).
\end{equation}
The strength of the random force $g_{ij}$ is related to the friction tensor $\gamma_{ij}$ by the classical Einstein relation,
\bel{einstein}
\sum_{k} g_{ik}g_{jk}=\gamma_{ij}T.
\end{equation}

In principle, the inertia and friction tensors may contain the
shell effects too. To account for these effects one could
consider the microscopic transport coefficients  calculated, for
example, within the linear response theory and local harmonic
approximation \cite{ivan97,hofm97,yama97}. It turns out that
friction tensor calculated by the microscopic model is
temperature dependent and much smaller than that calculated by the
macroscopic model at low temperature. As it follows from the
present calculations, the MDFF does not depend much on the
magnitude of the friction tensor \cite{arit13}. So, in present
work we use the macroscopic friction and inertia tensors. The
investigation of the role of shell effects in the collective
inertia and friction coefficients will be the subject of future
work.

The temperature $T$ is related to the intrinsic excitation energy of the composite system as $E_{int} =
\bar aT^2$, where $E_{\rm int}$ is calculated at each step of a trajectory calculation as
\begin{equation}\label{eint}
E_{\rm int}=E^{*}-\frac{1}{2}\left(m^{-1}\right)_{ij}p_{i}p_{j}-V(q, T=0).
\end{equation}
The excitation energy of compound nucleus $E^{*}$ is given by $E^{*}=E_{\rm cm}-Q$, where $Q$ denotes the $Q$-value of the reaction.

The calculation starts from the ground state, which is located at
$\bar z_0 = 0.0, \delta= 0.2, \alpha= 0.0 $. For the initial
distribution of collective velocities  we assume that at the
initial moment the collective kinetic energy of the system is zero. Very
soon after the start of  calculation the state of the system
becomes close to the statistical equilibrium. Such initial
conditions are very close to that used in earlier dynamical
calculations \cite{karp01}.

The fission events are determined in our calculations by classification of different
trajectories in the deformation space.
Fission from a compound nucleus is defined as the case that
a trajectory overcomes the scission point on the potential energy surface. The scission point is assumed here to be given by the configuration with zero neck radius.
%%%%%%%%%%%%%%%%%%%%%%%%%%%%%%%%%%%%%%%%%%%%%%%%%%%%%%%%%%%%%%%%%%%%%%%%%%%%%%%%%%%%
\begin{figure}[ht]
\includegraphics[width=.4\textwidth]{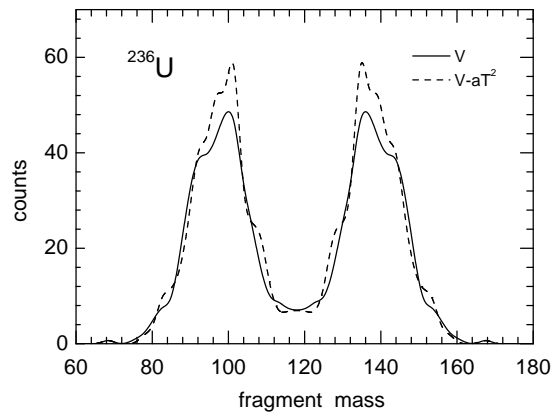}
  \caption{The comparison of the mass distribution of fission  fragments calculated with (dash line) and without ( solid line) account of $\bar a T^2$ term in \protect\req{lange1}. }
  \label{at2}
\end{figure}
%%%%%%%%%%%%%%%%%%%%%%%%%%%%%%%%%%%%%%%%%%%%%%%%%%%%%%%%%%%%%%%%%%%%%%%%%%%%%%%%%%%%%

In Fig. \ref{at2} we demonstrate the role of $\bar a T^2$ in the
potential energy in Langevin equations \req{lange1}. As it is
seen from the figure the use of the energy instead of free energy
in \req{lange1} does not influence much the MDFF of $^{236}$U at
$E^{*}=20$ MeV. So, in all calculations reported below the term $\bar a T^2$ in the
potential energy in Langevin equations \req{lange1} was neglected.
%%%%%%%%%%%%%%%%%%%%%%%%%%%%%%%%%%%%%%%%%%%%%%%%%%%%%%%%%%%%%%%%%%%%%%%%%%%%%
\section{The origin of the mass asymmetric fission}
%%%%%%%%%%%%%%%%%%%%%%%%%%%%%%%%%%%%%%%%%%%%%%%%%%%%%%%%%%%%%%%%%%%%%%%%%%%%%

In our previous study \cite{arit13}, we investigated the fission of $^{236}$U at the excitation energy $E^{*}=20$ MeV and have calculated the MDFF, which turned out to agree rather well with  experimental data showing that the mass-asymmetric component of the distribution of fission fragments is dominant.
In the present paper we try to clarify the origin
of the mass-asymmetric fission events of $^{236}$U at low excitation energy  by analyzing  the dynamical trajectories in our model.

To understand the  behavior of trajectories and their
contribution to the MDFF, we first consider the simple case when
the shell effects are neglected and only the liquid drop energy
$V_{\rm LD}$ contributes to the potential energy surface (PES).

\begin{figure}[ht]
\includegraphics[width=.48\textwidth]{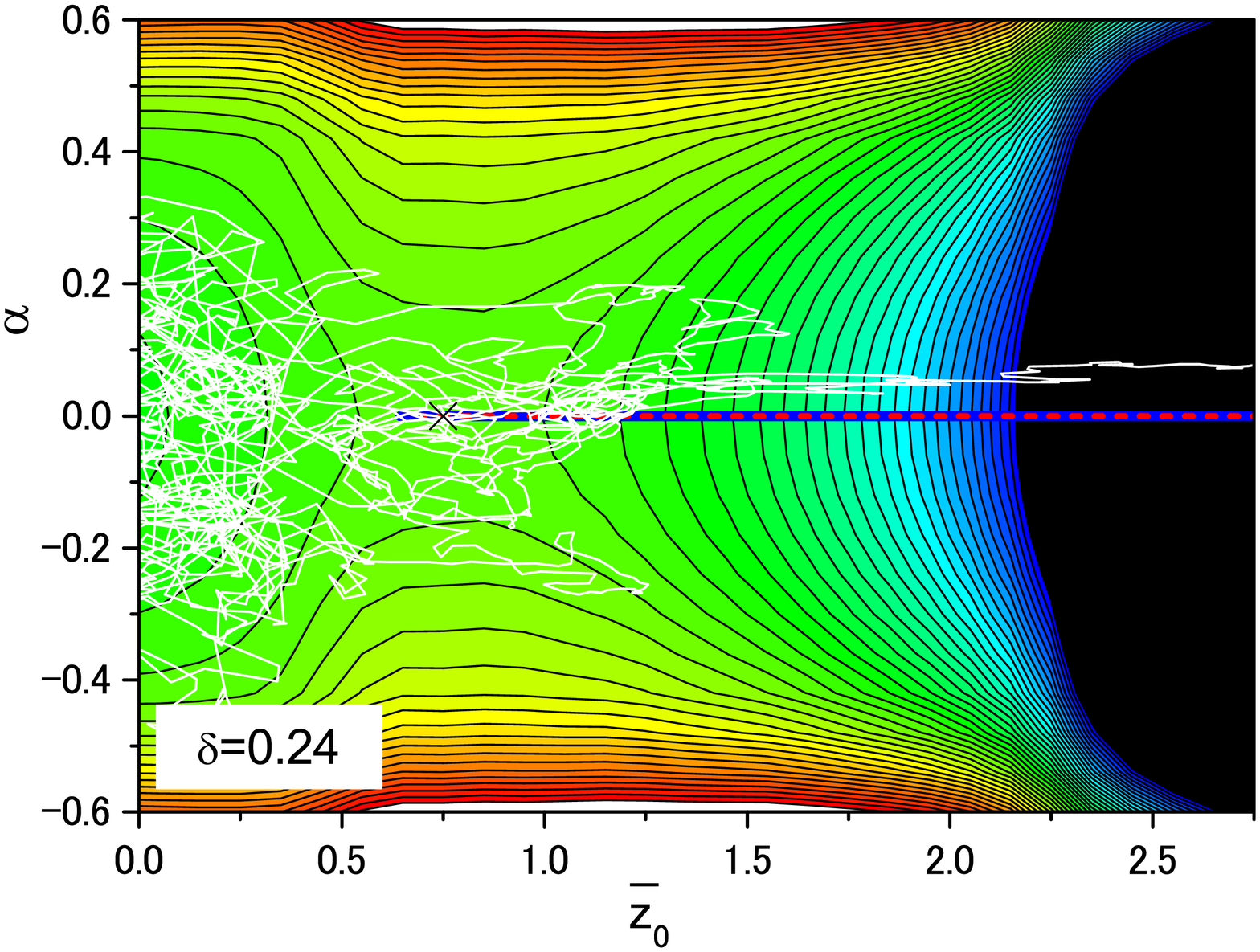}
\includegraphics[width=.48\textwidth]{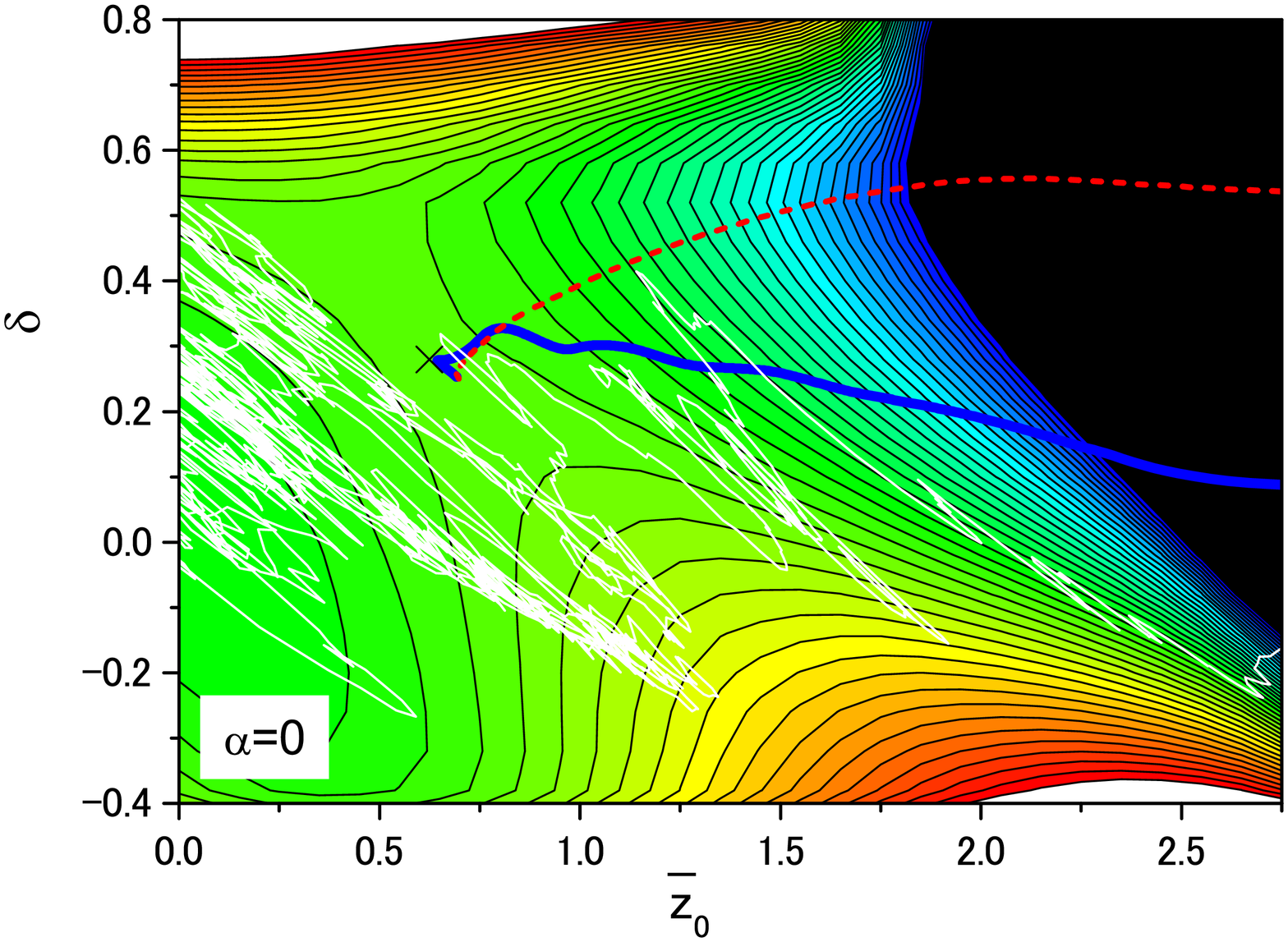}
\caption{The examples of trajectories of the fission process of $^{236}$U at $E^{*}=20 \mev$   projected onto the $\bar z_0$-$\alpha$  plane of $V_{\rm LD}$
$\delta=0.24$ (top) and on the $\bar z_0$-$\delta$ plane of $V_{\rm LD}$ at $\alpha=0$ (bottom).
The trajectories calculated without random force start at the saddle point.}
\label{fig2}
\end{figure}

Fig. \ref{fig2} shows the trajectories calculated with
 and without account of random force. To demonstrate
the relation to the potential energy the trajectories are placed
on the potential energy surface $V_{\rm LD}$ projected onto the
$\bar z_0$-$\alpha$ plane (top) and onto the $\bar z_0$-$\delta$ plane (bottom).

The trajectories calculated without account of random force
(heavy solid blue line) start at the saddle point deformation and
move down to the separation region along the potential slope due
to the drift force $-{\partial F}/{\partial q_{i}}$ in
Eq.~(\ref{lange1}). However, the trajectory does not move along
the line of steepest descent. This is a dynamical effect due to
the coupling between $\bar z_0$- and $\delta$-degrees of freedom
and the fact that the $\bar z_0 \bar z_0$- and
$\delta\delta$-components of mass tensor are very different. When
the nondiagonal components of mass tensor are neglected (dash red
line) the trajectory moves, indeed, along the line of steepest
descent, see bottom part of Fig. \ref{fig2}.
%%%%%%%%%%%%%%%%%%%%%%%%%%%%%%%%%%%%%%%%%%%%%%%%%%%%%%%%%%%%%%%%%%%%%%%%%%%%%%%%%
%%%%%%%%%%%%%%%%%%%%%%%%%%%%%%%%%%%%%%%%%%%%%%%%%%%%%%%%%%%%%%%%%%%%%%%%%%%%%%%

In case that the random force is  taken into account, the trajectories (thin solid white lines) show the oscillations in the direction of -45$^\circ$ on the
$\bar z_0$-$\delta$ plane.

We have analyzed the reason for this special direction and found
out that it originates from the properties of  the friction
tensor, mainly the nondiagonal terms, via the Einstein relation.
The detailed explanation will be given separately in a
forthcoming paper.

\begin{figure}[ht]
\includegraphics[width=.48\textwidth]{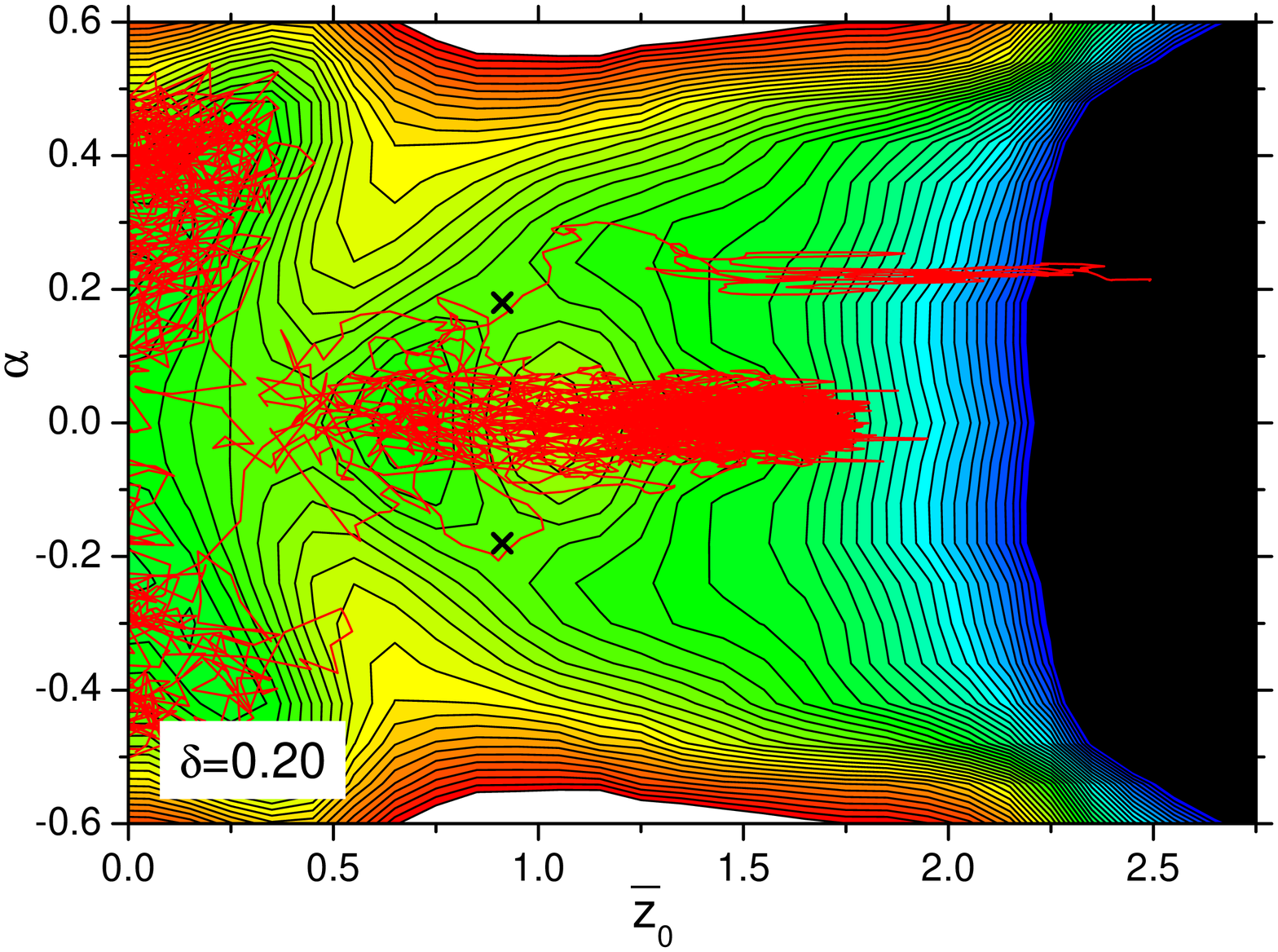}
\includegraphics[width=.48\textwidth]{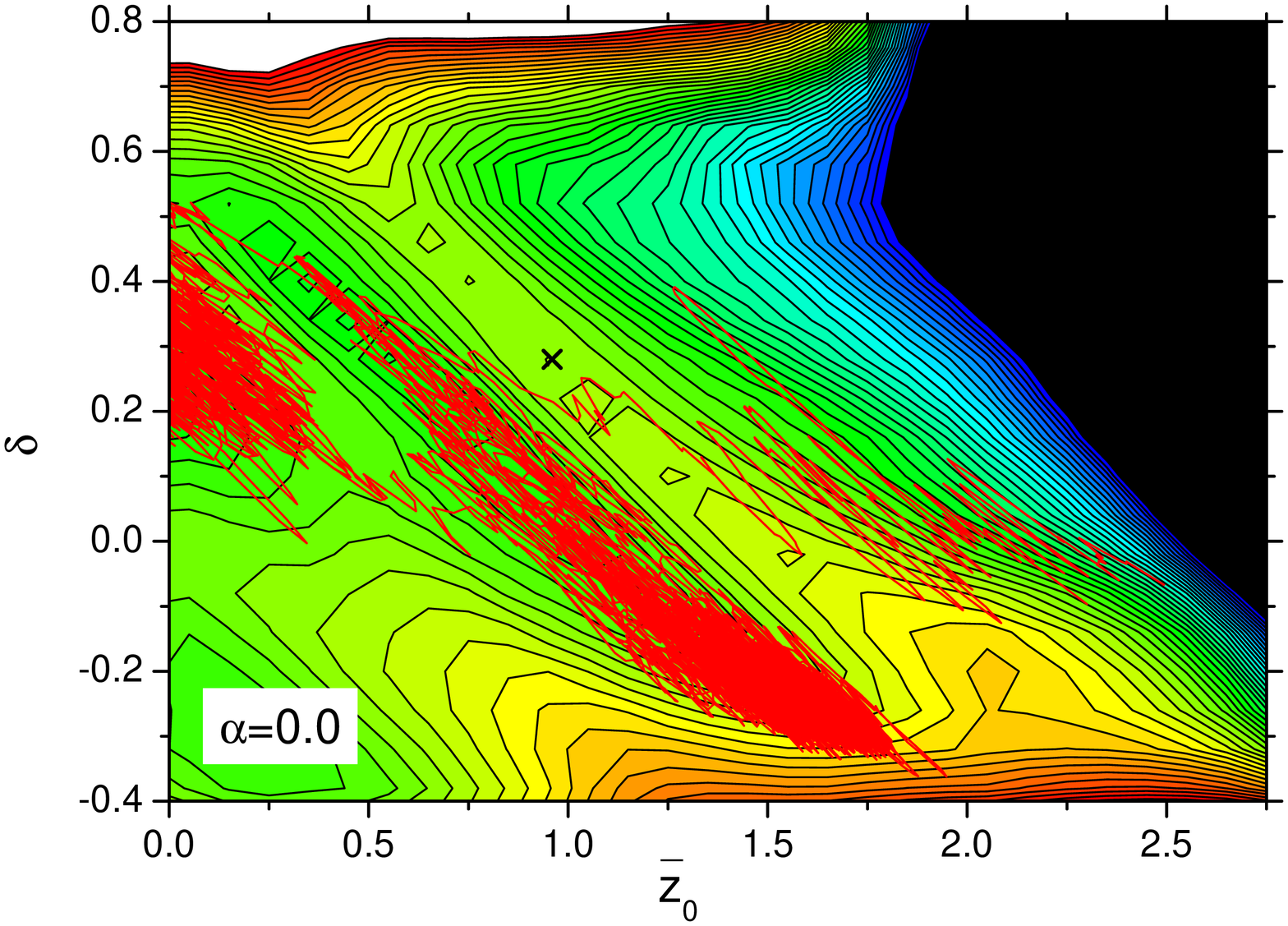}
\caption{Sample trajectory projected onto the $\bar z_0$-$\alpha$
plane at $\delta = 0.2$ (top) and the $\bar z_0$-$\delta$ plane $\alpha = 0.0$ (bottom) of
$V_{\rm LD}+E^{0}_{\rm shell}$ with $\epsilon = 0.35$ for $^{236}$U.
The trajectory starts
at the ground state $\{\bar z_0, \delta, \alpha\} =\{0.0, 0.2, 0.0\}$ at $E^{*}=20$ MeV.
The fission saddle points are indicated by the symbol $\times$. The scission lines are
denoted by the white lines}
\label{fig3}
\end{figure}

Fig.~\ref{fig3} shows the sample trajectories to the
mass-asymmetric fission region calculated with account both of
shell effects and random force, placed on the potential energy
surface $V_{\rm LD}+E^{0}_{\rm shell}$ and projected onto the
$\bar z_0$-$\alpha$ plane (top) and $\bar z_0$-$\delta$ plane (bottom).
Similarly to the calculation in reference \cite{arit13}, the trajectories start
at $\{\bar z_0,\delta,\alpha\}=\{0.0,0.2,0.0\}$, which corresponds to
the ground state of the potential energy surface. One can see,
that trajectories remain at the ground state (the first minimum) and the second
pocket for quite a long time. They even reach large $\bar z_0$ values of  $\bar z_0
=1.5 - 1.75$.  However, they do not move along straight line to
the separation region on the mass-symmetric fission path. Instead, the trajectories pass through the saddle points before moving to the scission region. It is
seen that the trajectories leading to mass-asymmetric fission
escape from the region around $\{\bar z_0,\alpha\} \sim \{0.8, \pm 0.2
\}$.
In Fig.~\ref{fig3}, the fission saddle points are indicated by the symbol $\times$.

It is seen that the mass-asymmetric fission originates from the trajectories that overcome the fission saddle points located at the mass asymmetry corresponding to the position of the peak of the MDFF, where $A \sim 140$.
One can note also that the trajectory on the $z$-$\delta$ plane fluctuates in the direction of -45$^\circ$, as shown in Figs.~\ref{fig2}
and \ref{fig3}.
Even after overcoming the fission saddle point, such oscillations are observed up to the scission point.

The reason of these fluctuations may be understood in a following way.

\begin{figure}[ht]
\includegraphics[width=.4\textwidth]{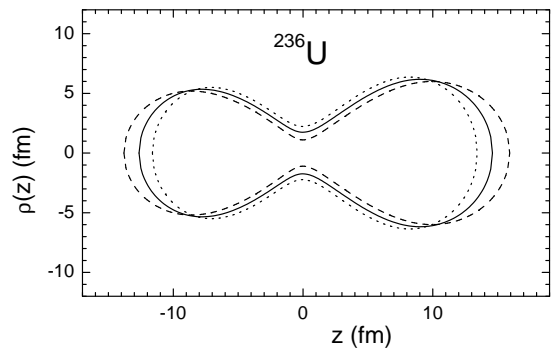}
  \caption{Nuclear shapes around the scission point of $^{236}$U.
The dot, solid and dash line corresponds to the nuclear shape at $\{\bar z_0,\delta,\alpha\}=\{2.5, -0.2, 0.2\}$, $\{2.5, 0, 0.2\}\, \text{and}\, \{2.5, 0.2, 0.2\}$.}
\label{fig4}
\end{figure}

We know from the analysis in \cite{cahaivta} that the mass distribution of fission fragments of $^{236}$U can be described in terms of {\it three} fission modes. Two of them are mass-asymmetric, the so called standard and super-short modes. They differ by the elongation of the shape in the scission region.
In principle, there should be two mass-asymmetric valleys in the potential energy surface   shown in the bottom part of Fig. \ref{fig3}. In principle, by fluctuations the system can jump from one valley to other. We do not see the two valleys in PES, but the landscape of potential energy is very flat in the direction smaller elongation $\Longleftrightarrow$ larger elongation (-45$^\circ$). Within TCSM  smaller elongation $\Longleftrightarrow$ larger elongation corresponds to positive $\delta$ $\Longleftrightarrow$ negative $\delta$, please, see the demonstration in Fig. \ref{fig4}.
%%%%%%%%%%%%%%%%%%%%%%%%%%%%%%%%%%%%%%%%%%%%%%%%%%%%%%%%%%%%%%%%%%%%%%%%%%%%%%%%%
%%%%%%%%%%%%%%%%%%%%%%%%%%%%%%%%%%%%%%%%%%%%%%%%%%%%%%%%%%%%%%%%%%%%%%%%%%%%%%%%%

If the potential PES is flat even small random force can cause the shape fluctuations of large amplitude. Something like this is observed in present calculations.

So, we would interpret the fluctuations on the way to the scission point observed in our calculations as the transitions between compact and elongated shapes, that both contribute to the mass distributions and TKE.
%%%%%%%%%%%%%%%%%%%%%%%%%%%%%%%%%%%%%%%%%%%%%%%%%%%%%%%%%%%%%%%%%%%%%%%%%%%%
%%%%%%%%%%%%%%%%%%%%%%%%%%%%%%%%%%%%%%%%%%%%%%%%%%%%%%%%%%%%%%%%%%%%%%%%%%%
%%%%%%%%%%%%%%%%%%%%%%%%%%%%%%%%%%%%%%%%%%%%%%%%%%%%%%%%%%%%%%%%%%%%%%%%%%%%%%%%%%
\section{The width of peak in MDFF}
%%%%%%%%%%%%%%%%%%%%%%%%%%%%%%%%%%%%%%%%%%%%%%%%%%%%%%%%%%%%%%%%%%%%%%%%%%%%%%%%%%%
The oscillations of the trajectories is a very important feature
of the fission dynamics. After overcoming the fission saddle point, as
shown in Fig.~\ref{fig3}, the trajectories fluctuate frequently
and move down the potential slope step by step. The direction of
the oscillation is neither parallel nor perpendicular to the
contour lines of the potential energy surface. The trajectories
climb and descend the potential slope in the result of the random
force and drift force, respectively. Correspondingly,  the
nuclear shape fluctuates around some average value as
demonstrated in Fig.~\ref{fig4}. The edges of nuclear shape with
$\delta < 0$ (dot line in Fig.~\ref{fig4}) are oblate, the
curvature of the edge sides is smaller as compared with spherical
shape ($\delta = 0$). In spite of the negative $\delta$ parameter,
the total shape of heavy and light fragments is, of course,
prolate (quadruple moments of left and right parts of fissioning
nucleus close to the scission point are positive).

Due to fluctuations, the nuclear fission does not occur with
continuous stretching, as exhibited by starch syrup. Beyond the
saddle and around the scission point the vibration of the length
and breadth of the fissioning fragments takes place until the
nucleus is split suddenly into two pieces by a strong vibration of
the length ($-\delta$ direction), which reduces the density in
the neck region \cite{iwam13}. We can conclude that the width of
the peaks of the MDFF is determined by such fluctuations near the
scission point. Since the calculated in \cite{arit13} MDFF is in
a good agreement with the experimental data, it supports our
conclusion that the vibration of the nuclear shape is essential
to describe nuclear fission correctly.
%%%%%%%%%%%%%%%%%%%%%%%%%%%%%%%%%%%%%%%%%%%%%%%%%%%%%%%%%%%%%%%%%%%%%%%%%%%%%%%%%%%%%%%%%%%%%%%
\section{Nuclear shape at scission point and the Total Kinetic Energy}
%%%%%%%%%%%%%%%%%%%%%%%%%%%%%%%%%%%%%%%%%%%%%%%%%%%%%%%%%%%%%%%%%%%%%%%%%%%%%%%%%%%%%%%%%%%%%%%
In Sec.~IV, we pointed out that the nuclear shape with a negative
$\delta$, particularly around the scission point, is very important in fission dynamics. The motion of trajectories in the
negative $\delta$ direction driven by the random force leads to
the splitting into fragments. The examples of nuclear shapes
around the scission point are presented in Fig.~\ref{fig4}. The
shape denoted by the dot line is close to that obtained using the
statistical scission model in Fig.~5 of reference \cite{fong56}.

\begin{figure}[ht]
\includegraphics[width=.48\textwidth]{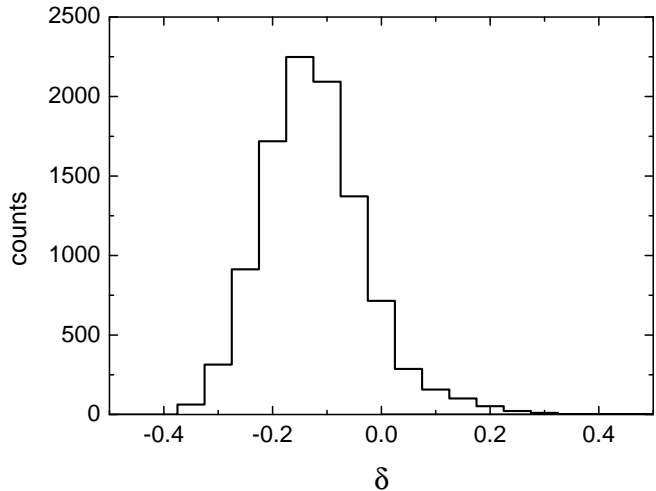}
  \caption{The calculated distribution of fission events in the deformation parameter $\delta$ at the scission point for the fission of $^{236}$U at $E^{*}=20$ MeV.}
\label{fig_a12}
\end{figure}

The distribution of the deformation parameter $\delta$ at the
scission point for the fission of $^{236}$U at $E^{*}=20$ MeV is
shown in Fig.~\ref{fig_a12}.
%%%%%%%%%%%%%%%%%%%%%%%%%%%%%%%%%%%%%%%%%%%%%%%%%%%%%%%%%%%%%%%%%%%%%%%%%%%%%
%%%%%%%%%%%%%%%%%%%%%%%%%%%%%%%%%%%%%%%%%%%%%%%%%%%%%%%%%%%%%%%%%%%%%%%%%%%%%%%%%%%%%%%%%
One can see that the contribution of fragments with negative values of $\delta$ is dominant.

To clarify the configuration at the scission point, we investigate the TKE of the fission fragments.
The scission configuration is defined as the shape with neck radius equal to zero \cite{arit13}, and the TKE is assumed to be given by
\begin{equation}
TKE=V_{Coul}+E_{pre}\,,
\end{equation}
where $V_{Coul}$ and $E_{pre}$ are the Coulomb repulsion energy
of point charges of fragments and the pre-scission kinetic energy.
The $V_{Coul}$ is defined as
$V_{Coul}=Z_{1}Z_{2}e^{2}/D$, where $Z_1$ and $Z_2$ are the charges
of each fragment, and $D$ is the distance between centers of mass
of left and right parts of nucleus at the scission point. We do not assume that
the distance between centers of mass is that of the nascent
fragments, like it is done in the statistical models in order to
calculate the TKE.

The prescision kinetic energy $E_{pre}$ is the kinetic energy
\begin{equation}\label{ekin}
E_{kin}=\frac{1}{2}\left(m^{-1}\right)_{ij}p_{i}p_{j}
\end{equation}
calculated at the scission point. The average of $E_{pre}$ over all fission events is equal to 7.03 MeV. So, the main contribution to the total kinetic energy comes from the Coulomb repulsion of fission fragments.

For the average value of TKE of the fission fragments $\langle TKE \rangle$ of $^{236}$U at $E^{*}=20$ MeV we obtained $\langle TKE \rangle = 169.5$ MeV, what is in agreement with the experimental data (168.2 $\sim$ 171 MeV) \cite{vand73}.
%Also, the dependence of $\langle TKE \rangle$ on the mass number of the fission fragments was calculated, which was shown in Fig.~7
%of reference \cite{arit13}.
Because of this agreement with the experimental data for the TKE, we conclude that the configuration
at the scission point is compact, such as that shown by the dot line in Fig.~\ref{fig4}.

\begin{figure}[ht]
\includegraphics[width=.48\textwidth]{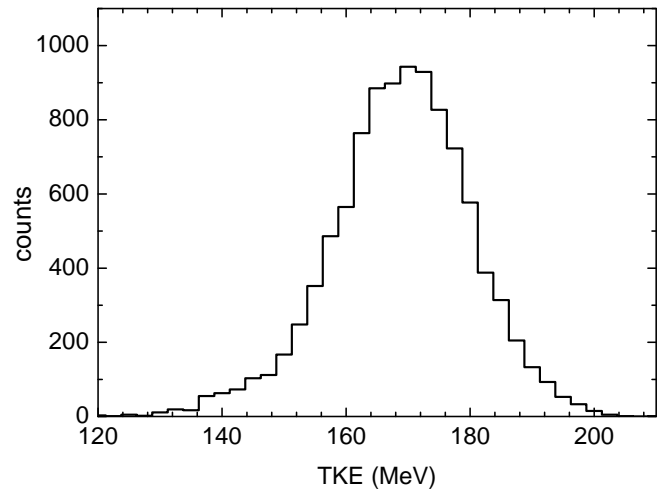}
\caption{The TKE distribution of the fission events of $^{236}$U at $E^{*}=20$ MeV.}
\label{fig_a13}
\end{figure}

The TKE distribution of the fission fragments of this system is shown in Fig.~\ref{fig_a13}.
%%%%%%%%%%%%%%%%%%%%%%%%%%%%%%%%%%%%%%%%%%%%%%%%%%%%%%%%%%%%%%%%%%%%%%%%%%%%%%%%%%%%%%%%
%%%%%%%%%%%%%%%%%%%%%%%%%%%%%%%%%%%%%%%%%%%%%%%%%%%%%%%%%%%%%%%%%%%%%%%%%%%%%%%%%%%%%%%%
The distribution is approximately Gaussian.
%%%%%%%%%%%%%%%%%%%%%%%%%%%%%%%%%%%%%%%%%%%%%%%%%%%%%%%%%%%%%%%%%%%%%%%%%%%%%%%%%%%%
\begin{figure}[ht]
\includegraphics[width=.48\textwidth]{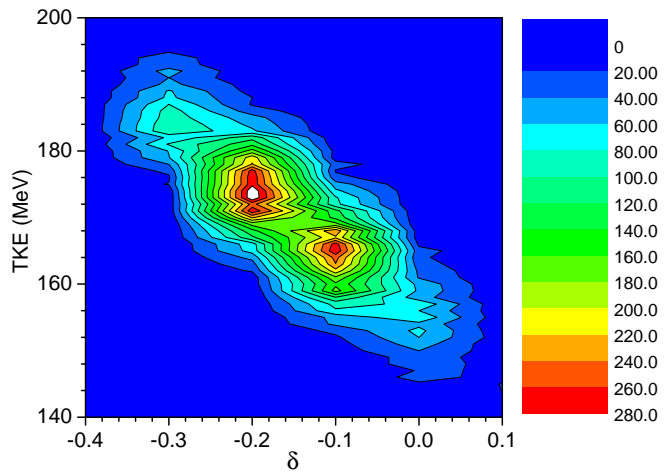}
  \caption{The distribution of fission events of $^{236}$U at $E^{*}=20$ MeV in the TKE and the parameter $\delta$.}
\label{fig7}
\end{figure}
%%%%%%%%%%%%%%%%%%%%%%%%%%%%%%%%%%%%%%%%%%%%%%%%%%%%%%%%%%%%%%%%%%%%%%%%%%%%%%%%%%%%

Fig. \ref{fig7} shows the distribution of fission events in the total kinetic energy and parameter $\delta$. From this figure one can see the correlation between the TKE and the value of parameter $\delta$ at the scission line.
The configuration with a negative $\delta$ corresponds to the compact shape. The TKE of such fragments is higher than that of fragments with a positive $\delta$.
The fissioning fragments with the compact configuration are dominant in this system.

%%%%%%%%%%%%%%%%%%%%%%%%%%%%%%%%%%%%%%%%%%%%%%%%%%%%%%%%%%%%%%%%%%%%%%%%%%%%%%%%%%%%
\section{Summary}
%%%%%%%%%%%%%%%%%%%%%%%%%%%%%%%%%%%%%%%%%%%%%%%%%%%%%%%%%%%%%%%%%%%%%%%%%%%%%%%%%%%%
In present paper we investigated the fission process at low excitation energy using the Langevin equations.
By analyzing the trajectories calculated within our model \cite{arit13},
we have clarified the contributions of the mass-asymmetric fission events of $^{236}$U at low excitation energy.
In this way we gave an explanation for the mass-asymmetric fission of $^{236}$U by the dynamical approach based on fluctuation-dissipation theorem.

The mass distribution of fission fragments of $^{236}$U at
$E^{*}=20$ MeV is mass-asymmetric. We have found out that
position of the peak is  related to the position of the saddle point,
which is defined mainly by the shell correction energy. In order
to escape from the potential pocket around the ground state or
the second minimum, almost all trajectories pass through the
fission saddle point and move to the mass-asymmetric fission
region. After overcoming the fission saddle points, the
trajectories fluctuate frequently due to the random force in
the Langevin equation and approach the scission point. The
fluctuation around the scission point determines the widths of
the peaks of the MDFF.

By analyzing the fission process and investigating the shape
evolution, we have found that the motion in the negative
$\delta$ direction around the scission point is essential for the
fission process. We stress that nuclear fission does not occur
with continuous stretching, such as that observed in starch syrup.
Rather, due to the shape
vibration of the length and breadth of the fissioning fragments,
the nucleus is suddenly split by a strong vibration in the
negative $\delta$ direction. Such a mechanism in fission dynamics
and the configuration with negative $\delta$ values at the
scission point are supported by the fact that the calculated MDFF
and TKE show  good agreement with the experimental data in
reference \cite{arit13}.

In addition, we pointed out that the
trajectories do not always move along the bottom of the
potential energy valley owing to the random force, nor fluctuate
around the trajectory without the random force (mean trajectory).
Although the analysis of the fission process using the static
potential energy surface gives a reasonable results in some cases, it is not
enough to describe the complicated dynamics of the fission process. In this paper,
we stress the importance of the dynamical treatment of the fission process.

%--------------------------------------------

As further study, we plan to improve the model by increasing the
number of variables, namely by introducing independent
deformation parameters $\delta_1$ and $\delta_2$ for each fragment. We have also to
consider the effects of nuclear structure on the transport
coefficients and the fact that Einstein relation that does not
hold true at low excitation energies \cite{ivan97,hofm97,yama97}.
Moreover, the neutron emission from the fissioning system and
from the fission fragments should be included in the model. With such
improvements of the model, we aim to diminish the differences
between the calculated MDFF and the experimental data.

\section*{Acknowledgments}

Present study is the results of ``Comprehensive study of
delayed-neutron yields for accurate evaluation of kinetic of
high-burn up reactors" entrusted to Tokyo Institute of Technology
by the Ministry of Education, Culture, Sports, Science and
Technology of Japan (MEXT). The authors are grateful to Dr.
A.~Iwamoto, Prof. M.~Ohta, Prof. T.~Wada, Dr. K.~Nishio,
Dr.~A.V.~Karpov and Prof.~V.I.~Zagrebaev for
their helpful suggestions and valuable discussions. Special
thanks are deserved to Mr.~K.~Hanabusa (HPC Systems Inc.)  for
his support to operate the high performance computer.

\section*{Appendix}

\begin{figure}[ht]
\includegraphics[width=.4\textwidth]{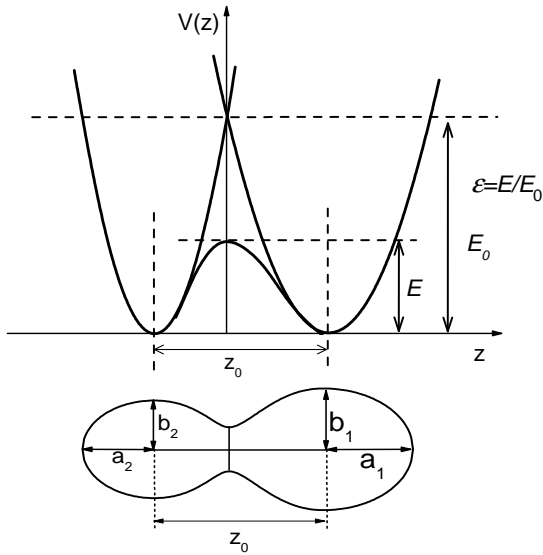}
\caption{The $z$-dependence of the potential $V(\rho, z)$ \protect\req{vofz} (top) and the  example of the equipotential surface of potential $V(\rho, z)$ (bottom). The neck parameter $\epsilon$ is defined as the ratio of the smoothed potential height $E$ at $z=0$ to the original one $E_{0}$.}
\label{def1}
\end{figure}

The momentum independent part $V(\rho, z)$ of the TCSM
Hamiltonian is formed by the two deformed oscillator potentials
joined smoothly by the fourth order polynomial in $z$, \bel{vofz}
V(\rho, z)=\left\{
\begin{array}{rl}
\frac{1}{2}m\omega_{z1}^2 \,(z-z_1)^2+\frac{1}{2}m\omega_{\rho1}^2 \,\rho^2,\, z\leq z_1\\
\frac{1}{2}m\omega_{z1}^2 \,(z-z_1)^2 f_1(z, z_1)\qquad\qquad\quad\quad\\
+\frac{1}{2}m\omega_{\rho1}^2 \,\rho^2 f_2(z, z_1),\quad\quad\,\,\,  z_1\leq z \leq 0\\
\frac{1}{2}m\omega_{z2}^2 \,(z-z_2)^2 f_1(z, z_2)\qquad\qquad\quad\quad\\
+\frac{1}{2}m\omega_{\rho2}^2 \,\rho^2 f_2(z, z_2),\quad\quad\,\,\,  0\leq z\leq z_2\\
\frac{1}{2}m\omega_{z2}^2 \,(z-z_2)^2+\frac{1}{2}m\omega_{\rho2}^2 \,\rho^2,
 z_2\leq z , \\
\end{array} \right.
\end{equation}
see Fig. \ref{def1}.
Here $m$ is a nucleon mass and $\omega_z,
\omega_{\rho}$ are the oscillator frequencies in $z$ and $\rho$
directions. The functions $f_1$ and $f_2$  are given by
\belar{f1f2}
f_1(z, z_i)&=&1+c_i(z-z_i)/z_i+d_i(z-z_i)^2/z_1^2\,,\nonumber\\
f_2(z, z_i)&=&1+g_i(z-z_i)^2/z_1^2\,.\qquad i=1,2.
\end{eqnarray}

The potential $V(\rho, z)$ is characterized by 12 parameters, see \req{vofz}.
After imposing condition that the parts of potential are joined
smoothly at $z=z_1, z=z_2$ and $z=0$ the number of independent parameters is reduced to five: the distance $z_0=z_2-z_1$
between centers of oscillator potentials, the mass asymmetry
$\alpha$, the deformation $\delta_1$ and $\delta_2$ of left and
right oscillator potentials and the neck parameter $\epsilon$. The
neck parameter $\epsilon$ is given by the ratio of the
potential height $E$ at $z=0$ to the value $E_{0}$ of left and right harmonic oscillator potentials at $z=0$ (which should be the same), see Fig~\ref{def1}.

All the parameters appearing in \req{vofz} are expressed in terms
of these 5 deformation parameters \belar{tcsm_params}
z_1&=&-z_0\,\omega_{z2}/(\omega_{z1}+\omega_{z2}),\,\,
z_2=z_0\,\omega_{z1}/(\omega_{z1}+\omega_{z2}),
\nonumber\\
c_1&=&c_2=2-4\eps,\quad d_1=d_2=1-3\eps\\
g_1&=&-\frac{\omega_{\rho1}^2-\omega_{\rho2}^2}{\omega_{\rho1}^2}\frac{\omega_{z2}}{\omega_{z1}+\omega_{z2}},\,\,
g_2=\frac{\omega_{\rho1}^2-\omega_{\rho2}^2}{\omega_{\rho2}^2}\frac{\omega_{z1}}{\omega_{z1}+\omega_{z2}}\nonumber
\end{eqnarray}

The ratio of oscillator frequencies
${\omega_{\rho}}/{\omega_{z}}$ or the ratio of semi-axes in
$\rho$ and $z$ directions are related to the deformation
parameters $\delta_i$  , \bel{deltai} \frac{\omega_{\rho
i}}{\omega_{zi}}= \frac{a_i}{b_i}=( 1-\frac{2}{3}\delta_i
)\Big\slash( 1+\frac{1}{3}\delta_i),
\end{equation}
see Fig. \ref{def1}. Notice that $\delta_i < 1.5$ since  $a_i > 0$ and $b_i > 0 $.
%%%%%%%%%%%%%%%%%%%%%%%%%%%%%%%%%%%%%%%%%%%%%%%%%%%%%%%%%%%%%%%%%%%%%%%%%%%%%%%%%%%%%%%%%
%%%%%%%%%%%%%%%%%%%%%%%%%%%%%%%%%%%%%%%%%%%%%%%%%%%%%%%%%%%%%%%%%%%%%%%%%%%%%%%%%%%%%%%%%

The ratio $\omega_{\rho1}/\omega_{\rho2}$ should be found numerically from the condition $(V_L-V_R)/(V_L+V_R)=\alpha$, were $V_L$ and $V_R$ are the volumes of left and right parts of nucleus. In TSCM parameterization the shape is divided in parts by the point $z=0$.

The parameter $\epsilon$ is defined in the same way as in Ref. \cite{maru72,zagr07}.
With $\epsilon < 1$, the surface of two fragments shows the smooth curve at the connecting point of
them. On the other hand, in the case of $\epsilon = 1$, the two fragments are connected with a sharp point like a top of cone.

We define the sharp surface shape of the fissioning nuclei $\rho(z)$ as that
given by the equipotential surfaces of potential $V(\rho, z)$,
i.e. by the equation \bel{rhoz} V(\rho(z), z)=V_0.
\end{equation}
The constant $V_0$ in \req{rhoz} is
found from the requirement that the volume inside equipotential
surface is equal to the volume of spherical nucleus.

The advantage of the two-center shell model parameterization is
that besides the commonly used in the theory of nuclear fission
degrees of freedom for the elongation, mass asymmetry and neck
radius it allows the independent variation of the deformation of
left and right parts of nucleus (fragments). Besides, this
parameterizaton describes both compact shapes and separated
fragments what makes it possible to describe fusion and fission
processes within the same shape parameterization.
%%%%%%%%%%%%%%%%%%%%%%%%%%%%%%%%%%%%%%%%%%%%%%%%%%%%%%%%%%%%%%%%%%%%%%%%%%%%%%%%%%%%
\begin{figure}[ht]
\includegraphics[width=.4\textwidth]{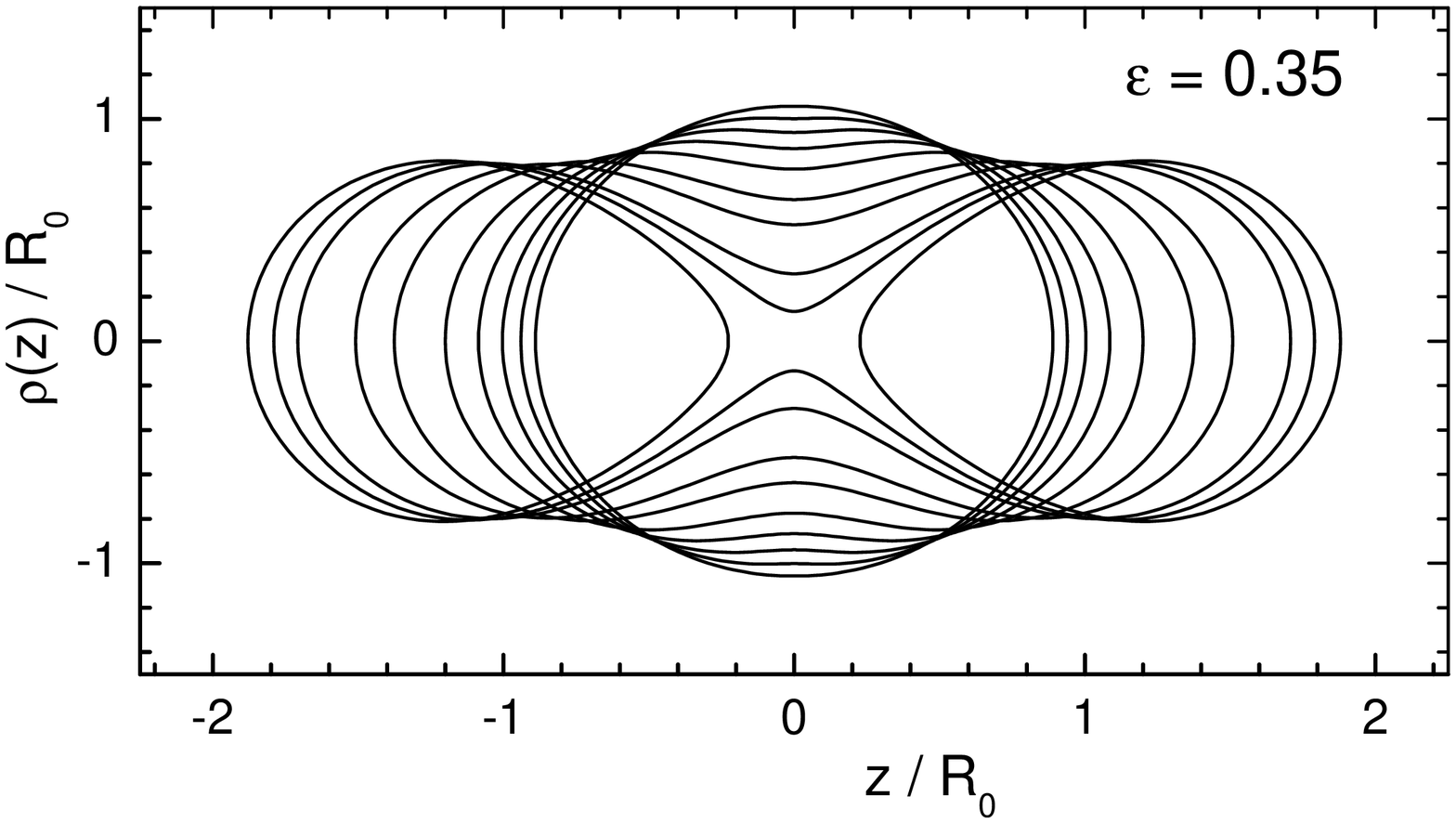}
  \caption{The examples of shapes \protect\req{rhoz} for few values of $z_0$ at fixed $\alpha=0,\, \delta_1=\delta_2=0, \,\epsilon=0.35$ .}
\label{shapes}
\end{figure}
%%%%%%%%%%%%%%%%%%%%%%%%%%%%%%%%%%%%%%%%%%%%%%%%%%%%%%%%%%%%%%%%%%%%%%%%%%%%%%%%%%%%%

Please, note that the neck radius depends on all 5 deformation
parameters. Even keeping neck parameter $\epsilon$ fixed one can
get a full variety of fission shapes from sphere to the two separated
fragments, see Fig. \ref{shapes}.

%-----------------------------------------------------------------

\end{document}